\documentclass[aps,nofootinbib,11pt]{revtex4-1}
\usepackage[letterpaper,hmargin=1in,vmargin=1in]{geometry}

\usepackage{graphicx,epstopdf,amsmath,amsfonts,amssymb,amsxtra,color,mathrsfs}

\usepackage{multirow,array}
\usepackage{cancel,bbold,extarrows}
\usepackage{hyperref}

\begin{document}

\title{SHENANIGANS AT THE BLACK HOLE HORIZON: PAIR CREATION OR BOULWARE ACCRETION?}

\author{Werner Israel\footnote{Honorary Fellow, Canadian Institute for Advanced Research}}
\email{israel@uvic.ca}

\affiliation{Department of Physics and Astronomy, University of Victoria, 
     Victoria, B.C., Canada, V8W 2Y2}

\date{\today}

\begin{abstract}
%\noindent
The current scenario of black hole evaporation holds that the Hawking energy flux $F$ is powered by pair creation at the horizon. However, pair creation produces entanglements, some of which must necessarily be broken before the black hole evaporates completely. That leads to loss of information and violation of unitarity.

In this paper, an alternative scenario is suggested that reproduces the essential features of Hawking evaporation, but does not invoke pair creation with its attendant problems. In this ``accreting Boulware'' scenario, a positive flux $F$ is still an outflux at infinity, but near the horizon it becomes an influx of negative energy. This negative energy flux (marginally) satisfies the Flanagan energy inequality.

Comments: References added, minor changes.

\end{abstract}
\maketitle

\section{Introduction}\label{intro}

In a famous 1974 paper \cite{Hawking1074}, Stephen Hawking announced his discovery that black holes evaporate by emitting thermal radiation. The source of this Hawking radiation is generally believed to be a process of pair creation at the horizon, with the positive-energy particle escaping outwards as radiation while its negative-energy partner falls inwards and reduces the mass of the black hole.

In recent years this scenario has come under increasing fire. Pair creation quantum-mechanically entangles the two partners of each pair across the horizon. If the black hole evaporates completely by this process, simple arguments \cite{Polchinski} show that some of these entanglements must get broken. This results in a violation of unitarity --- a pillar of quantum mechanics --- and attendant loss of information. This has motivated a number of radical and even desperate proposals that try to save the picture, most recently the idea of having a ``firewall'' at the horizon as a concrete entanglement-breaker \cite{Polchinski}.

The purpose of this paper is to suggest an alternative scenario which reproduces the broad features of the Hawking picture without recourse to pair creation and its attendant difficulties.

To begin, some heuristic remarks. The quantum theory of fields in curved space predicts that the gravitational field of a massive body such as a star polarizes the surrounding space, producing a partial (and normally very small) separation of the vacuum into positive and negative mass-energies.

The basic mechanism responsible for this effect is not really understood at present, but is necessarily of a quantum character. (Because of the principle of equivalence, classical gravity cannot discriminate between its effects on positive and negative masses, and cannot effect their separation.) I shall speak loosely of ``polarizing forces'' or ``Boulware forces''.

The effect is analogous to the well-know vacuum polarization of quantum electrodynamics. An electric field slightly splits the vacuum into positive and negative charges. In this case the opposite charges are still linked by their mutual electrostatic attraction, so the applied field induces a small dipole field. But this attraction is not present in the case of gravity. Again by the principle of equivalence, a negative mass repels (i.e., induces the same outward acceleration of) all other bodies, irrespective of their mass. The result is that the positive-mass partners tend to be driven outwards, leaving the star surrounded by a nett distribution of negative energy density --- a so-called Boulware state \cite{modelBH}.

Suppose now that our star is slowly contracting. As its surface gravity becomes stronger the energy density around it becomes steadily more negative. The excess positive energy is expelled by Boulware forces. Finally, the star reaches its gravitational radius, the point at which a black hole is about to form.

Up to now the Boulware state has supported itself against gravity by a pressure gradient. But now the gradient needed has becomes infinite. Pressure support fails and negative Boulware energy begins to drain into the hole.

Meanwhile, further out, polarization forces are at work rebuilding the Boulware state and replenishing negative energy. Because of the loss of negative energy inwards they must expel an equal amount of positive energy outwards to maintain the right balance. The overall result is a steady flux of Hawking energy outwards while an equal flux of negative energy is absorbed by the hole. This, in a nutshell, is the Boulware accretion scenario. Pair creation does not appear in this semiclassical description.

Could "Boulware forces" just be pair creation in disguise? In that case, Boulware accretion has not replaced pair creation, it has only transferred it from the horizon to the hole's exterior. However, it is arguable (insofar as a classical picture of the process can be trusted) that Boulware accretion cannot involve pair creation. Local observers would actually measure the energy of the infalling Boulware flux to be negative. This cannot be a flow of real particles.

This stands in contrast to the picture of pair creation at the horizon. Here, the entangled pair have nearly parallel momenta above and below the horizon. Local observers measure the energy of both to be positive. (True, the conserved ``Killing energy'' $E=-p_a\xi^a$ of the nether particle is negative. But this is not locally measured energy; it includes a negative ``potential energy'' contribution.)

In the following sections we flesh out these ideas in the context of a simple (1+1)-dimensional model.

Freewheeling talk about negative energy quite rightly raises concerns. We address this question in the last section. There it is shown that the negative energies involved here are of the innocuous kind, like those of the Casimir effect. They do not actually contravene any law of physics. We conclude with some speculative remarks about black hole complementarity.

\section{REDUCED SPHERICAL EINSTEIN THEORY}

For a spherical metric in the general form
\begin{equation}
\operatorname{d}\!s^2=g_{ab}\operatorname{d}\!x^a\operatorname{d}\!x^b+r^2(x^a)\operatorname{d}\!\Omega^2
\end{equation}
$(a,b=0,1)$ the 4-dimensional Einstein-Hilbert Lagrangian reduces to
\begin{equation}\label{L}
L=\frac{1}{4}r^2R+\frac{1}{2}(\nabla r)^2+L_\mathrm{mat},
\end{equation}
where $R$ is the 2-dimensional curvature scalar.

We shall adopt the same Lagrangian to define a ``spherical'' Einstein theory in (1+1) dimensions, with $r(x^a)$ now an auxiliary scalar (``dilaton'') field.

Define the scalar fields $f,m$ by
\begin{equation}\label{fm}
f=(\nabla r)^2=1-2m/r.
\end{equation}
Then \eqref{L} leads to the field equations
\begin{align}
\partial_am&=(T_a^b-\delta_a^bT_d^d)\partial_br\label{field_eqn_1}\\
r_{;ab} &=\bigl(\square r-\frac{m}{r^2}\bigr)g_{ab}-\frac{1}{r}T_{ab},\label{field_eqn_2}
\end{align}
which imply the conservation laws
\begin{equation}
T^b_{a;b}=0.
\end{equation}

For access to the future horizon it is useful to work with an advanced time co-ordinate $v$. A general 2-metric then takes the form
\begin{equation}\label{metric_vr}
\operatorname{d}\!s^2=2e^\psi\operatorname{d}\!v\operatorname{d}\!r-fe^{2\psi}\operatorname{d}\!v^2,
\end{equation}
and the field equations \eqref{field_eqn_1}, \eqref{field_eqn_2} become
\begin{equation}\label{field_eqn_3}
m_r=-T_v^v,\qquad m_v=T_v^r,\qquad\psi_r=\frac{1}{r}T_{rr}.
\end{equation}
(Subscripts on $m$ and $\psi$ indicate partial derivatives.)

The curvature scalar for metric \eqref{metric_vr} is
\begin{equation}\label{Rscalar}
R=-2e^{-\psi}\kappa_v,\qquad\kappa\equiv\frac{1}{2}e^{-\psi}(fe^{2\psi})_r+\psi_v,
\end{equation}
where $\kappa$ is closely related to the local surface gravity (redshifted acceleration of a static observer):
\begin{equation}
(-g_{00})^{\frac{1}{2}}a=\kappa-\frac{1}{2}\partial_v\ln(fe^{2\psi}).
\end{equation}

As source $T^a_b$ we consider a massless scalar field propagating on this classical background. Expectation values of all components $T^a_b$ can be obtained from the conservation laws and the 2-dimensional conformal anomaly \cite{modelBH}
\begin{equation}\label{anomaly}
T_a^a=\tilde hR,\qquad\tilde h\equiv\hbar/24\pi.
\end{equation}

By manipulation of the conservation laws and use of \eqref{anomaly} one arrives at
\begin{equation}\label{conserv1}
\partial_r\{fe^{2\psi}T_r^r+2e^\psi T_v^r+\tilde h(\kappa^2-2\partial_v\kappa)\}=0
\end{equation}
(details in Appendix A). This provides a convenient first integral of the conservation laws.

It is helpful to recast \eqref{conserv1} in terms of what stationary observers (world-lines $r=\text{const.}$) would actually measure. We shall focus on the hole's exterior (outside the apparent horizon), where $f>0$. Introduce an orthonormal pair of basis vectors
\begin{equation}
s_a=f^{-\frac{1}{2}}\partial_ar,\qquad t^b=(fe^{2\psi})^{-\frac{1}{2}}\partial x^b/\partial v
\end{equation}
which satisfy
\begin{equation}
s_as^a=-t^bt_b=1,\qquad s_at^a=0.
\end{equation}
Further, define energy flux $F$, pressure and density by
\begin{equation}
F=-T_b^as_at^b,\qquad P=T^a_bs_as^b,\qquad\rho=T^a_bt^bt_a.
\end{equation}
Then the stress components in \eqref{conserv1} and\eqref{field_eqn_3} can be written
\begin{equation}\label{T_comp}
T^r_r=P+F,\qquad T^r_v=-fe^\psi F,\qquad fT_{rr}=\rho+P+2F
\end{equation}
and \eqref{conserv1} takes the form
\begin{equation}\label{conserv2}
\partial_r\{fe^{2\psi}(P-F)+\tilde h(\kappa^2-2\partial_v\kappa)\}=0.
\end{equation}
Since there is no incoming flux from past lightlike infinity ($r=\infty,v=\text{const.}$), each term of \eqref{conserv2} vanishes there; \eqref{conserv2} integrates to
\begin{equation}\label{conserv_int}
fe^{2\psi}(P-F)=-\tilde h(\kappa^2-2\partial_v\kappa).
\end{equation}

A positive flux $F$ may signify an outflux of positive energy. Or it may signify an influx of negative energy, or a combination of the two. Our semi-classical theory is sphinx-like on this issue. Near infinity, it is most naturally interpreted as an outflux, since there is no external source. Near the horizon, the standard pair-creation scenario interprets positive $F$ as an outflux. But in the accreting Boulware scenario we shall interpret it as an influx of negative energy.

We work throughout with expectation values in a running Boulware ground state free of positive-frequency modes $e^{-i\omega v}$. This carries the great advantage (especially for the accreting Boulware scenario) of providing seamless coverage of the transition from contracting star to evaporating black hole; but also the inconvenience that the components $T_{ab}$ (and also $\rho$, $P$ and $F$) become very large (though not infinite) at the apparent horizon of an evaporating black hole, because they incorporate blue shifts. (That could be straightforwardly fixed by a Bogoliubov transformation to Kruskal modes and the Unruh state. But we shall not do this here.)

\section{RADIATION/ACCRETION AT THE HORIZON}

Equations (\ref{Rscalar}b), \eqref{fm}, (\ref{field_eqn_3}a,c) and \eqref{T_comp} give
\begin{equation}
P+F=rK-\frac{m}{r}-re^{-\psi}\psi_v\notag
\end{equation}
where $K=\kappa e^{-\psi}$. Substituting this into \eqref{conserv_int} then yields a quadratic equation for $K$:
\begin{equation}\label{Ksq}
\tilde hK^2+rfK-A=0
\end{equation}
where
\begin{equation}
A=(2F+\frac{m}{r}+re^{-\psi}\psi_v)f+2\tilde he^{-2\psi}\kappa_v
\end{equation}

Letting $f\rightarrow0$ in \eqref{Ksq}, and bearing in mind that $fF$ is nonzero in this limit because of the blueshift in the flux $F$ near the horizon, we find
\begin{equation}\label{fF}
fe^{2\psi}F\rvert_{f=0}=\frac{1}{2}\tilde h(\kappa_0^2-2\partial_v\kappa_0)
\end{equation}
where $\kappa_0(v)\equiv\kappa\rvert_{f=0}$ is the surface gravity of the contracting apparent horizon.

Formally, the first term on the right-hand side is the standard expression for Hawking flux in (1+1) dimensions, corresponding to a temperature
\begin{equation}
T_H=\hbar\kappa_0/2\pi.
\end{equation}
The second term incorporates the Kraus-Wilczek effect \cite{Wilczek}, the small change of  surface gravity that accompanies passage of one quantum through the horizon.

In the pair-creation scenario, the left-hand side is an outflux of positive energy generated by pair creation at the horizon; in the accreting Boulware scenario, it is negative energy draining into the hole from the surrounding Boulware state.

For slow evaporation, the left-hand side is nearly constant along outgoing light rays, as can be seen from
\begin{equation}
\frac{\operatorname{d}}{\operatorname{d}\!r}(fe^\psi F)=\frac{4}{r}e^\psi F^2,
\end{equation}
which follows from the conservation laws. Thus, \eqref{fF} is also nearly the flux at infinity:
\begin{equation}
F(r\rightarrow\infty)\approx\frac{1}{2}\tilde h\kappa^2_0.
\end{equation}
In both scenarios this is an outflux. And it follows from standard (e.g., thermo field dynamical \cite{thermofield}) arguments that it has a (nearly) thermal spectrum. For external observers the two scenarios are indistinguishable.

The $(-\text{in})\leftrightarrow(+\text{out})$ invariance of $F$ in
\begin{equation}
rf\psi_r=\rho+P+2F
\end{equation}
means that the choice of $F$ that ensures horizon regularity in the Boulware accretion scenario also does the job for the pair creation scenario. At this level of description, the mathematics makes no distinction whatever between these two very different pictures.

\section{CONCLUDING REMARKS}

A scheme involving negative energy should not normally expect a warm welcome in respectable physics circles. Yet quantum theory does not completely rule out negative energies; a well-known example is the Casimir effect. What it does do is to place strong restrictions in the form of ``energy inequalities'' \cite{Ford}, on how negative the energy can get. In (1+1) dimensions an elegant inequality due to Flanagan \cite{Flanagan}, constrains fluxes of negative energy. In Appendix B it is shown that Boulware energy satisfies (marginally and rather remarkably) the Flanagan constraint. Thus, Boulware accretion does not contravene any law of physics.

We thus have two very different pictures of what is happening near the horizon. But only one of these does not run foul of unitarity violation. Boulware accretion seems to be the more realistic picture.

Of course we are stretching classical pictures of quantum phenomena well beyond their expected range of validity. It may very well be that there is a kind of complementarity \cite{Susskind,War} at work here, and that, near the horizon, the two images will blur and merge into something not easily visualizable. The present difficulties with unitarity would then stand revealed as merely a case of grabbing hold of the wrong end of the complementary stick.

Note added: My thanks to Jim Bardeen and to Yasunori Nomura for useful comments, and for drawing my attention to their analyses of black hole evaporation \cite{Bardeen,Nomura} which have points in common with the accreting Boulware scenario.

\section*{Acknowledgements}

Warm thanks to Dr Xun Wang for stimulating discussions over the years, and for help with the manuscript.

\newpage

\section*{Appendix A: Derivation of (\ref{conserv1})}

If the two conservation laws $T^a_{b;a}=0$ are written out, and terms in $\partial_vf$ are eliminated between them, the result can be cast in the form
\begin{equation}
fe^\psi\partial_rT^r_r+2e^{-\psi}\partial_r(e^\psi T^r_v)+(T^r_r-\frac{1}{2}T^a_a)e^{-\psi}\partial_r(fe^{2\psi})+\partial_vT^a_a=0.
\end{equation}
Inserting the trace anomaly \eqref{anomaly} for $T_a^a$ and the expression \eqref{Rscalar} for $R$ into this equation leads, after some manipulations, to \eqref{conserv1}.

\section*{Appendix B: Flanagan's Inequality}

An observer (world-line $\mathscr{L}$, arbitrary 2-velocity $u^a$) moves through a stress-energy distribution $T_{ab}$ in 2-dimensional spacetime (curvature scalar $R$). Flanagan's inequality constrains the energy $\mathscr{E}$ encountered by this observer:
\begin{align}
\mathscr{E}(\mathscr{L},\gamma)&\equiv\int_\mathscr{L}\operatorname{d}\!\tau \gamma(\tau)T_{ab}u^au^b\geq\mathscr{E}_\mathrm{min}(\mathscr{L},\gamma)\\
\mathscr{E}_\mathrm{min}(\mathscr{L},\gamma)&=-\frac{\hbar}{24\pi}\int\operatorname{d}\!\tau\Bigl\{\Bigl(\frac{\dot\gamma}{\gamma}\Bigr)^2+\gamma a^2+\gamma R\Bigr\}.
\end{align}
Here $\gamma(\tau)$ is an arbitrary sampling function, satisfying
\begin{equation}
\gamma(\tau)\geq0,\qquad\int_{\mathscr L}\gamma(\tau)\operatorname{d}\!\tau=1.
\end{equation}

For the static Boulware state
\begin{equation}
\rho=P-\tilde hR=-\tilde h\Bigl(\frac{\kappa^2}{fe^{2\psi}}+R\Bigr)=-\tilde h(a^2+R),
\end{equation}
where $a$ the acceleration of a static observer. This marginally satisfies the Flanagan inequality (set $\gamma=1$).

\end{document}